# Simultaneous Measurement of Polarization and Excitation-Emission Spectrum of Suspended Particles in Water

Thesis Submitted to

**Tsinghua University**

in partial fulfillment of the requirement

for the professional degree of

by

**Liu Tong**

Thesis Supervisor : Associate Professor Liao Ran

Keywords: Multi-color, Polarization, EEM, Suspended single particle, Scattered light, Machine learning

Abstract

   The detection of suspended single particles (SSPs) in water is a crucial element for monitoring water ecosystems. Although there are a variety of sensory methods, it is still an unrealized goal to combine multi-wavelength polarized light scattering and fluorescence excitation-emission matrix (EEM) to characterize and classify SSP in water, which can deeply characterize and classify SSPs with different properties. In this paper, we propose an efficient optical method that can simultaneously obtain the SSP's Stokes parameters of four wavelengths and the fluorescence excitation-emission matrix (EEM) within 0.125μs at 120° scattering angle in water. We use Mie scattering simulation to validate our method and obtain the optimal parameters for the experiments. We demonstrate the high performance of our method in terms of sensitivity and accuracy through experiments involving microalgae, microplastics and sediments, as well as control experiments under different illumination conditions and classification experiments of harmful microalgae by using a multi-layer perceptron (MLP) to classify underwater suspended particles and harmful microalgae based on our method, and show that our method achieves better results than using single-wavelength light. Our method paves the way for the detection of SSPs and polarization research in marine science.

## 1    Introduction

   Suspended single particles (SSPs) such as microalgae, microplastic particles and sediments play an important role in the water environment[1-5]. For example, many microalgae, just like *Chlorella minutissima*, *Scenedesmus obliquus* and *Dunaliella tertiolecta*, are beneficial for the marine ecosystem[6], as they fix particulate carbon at the bottom of the food chain and provide energy for other organisms. However, some microalgae are harmful, as they can cause harmful algal blooms (HABs) that pose a serious threat to marine life and human health[7]. Moreover, with the increased frequency of red tide, the dominant species of phytoplankton has also changed[8].

   Therefore, *in-situ* detection of harmful microalgae is crucial to marine environment protection[9-11]. Another problem that affects the water environment is microplastic pollution, which has adverse impacts on marine economy and human health. For instance, marine plastic pollution, represented by plastic debris dumps, is an urgent global challenge that affects humans through the food chain[12], besides, there is still great room for development of optical detection methods for microplastics[13]. Furthermore, by studying the sediment deposition process, we can better understand the environment of river delta[14]. In summary, developing efficient optical methods for measuring SSPs in-situ is of great significance for marine science.

There are several optical detection methods for SSP in water, such as optical microscopy, Raman spectroscopy, Fourier transform infrared spectroscopy (FTIR), fluorescence excitation-emission matrix (EEM), and polarized light scattering measurement. Methods' inherent advantages often bring disadvantages. Optical microscopy is easy to operate, but it relies on experts' visual identification and experience, and it often requires sample preparation, extraction, fixation, or staining, which can be time-consuming and costly. Moreover, optical microscopes cannot perform underwater in-situ measurements[15-18]. Raman spectroscopy can identify the type of particles by using the Raman spectral characteristics of scattered light. This method has high sensitivity and specificity, and does not require sample treatment, allowing for rapid and accurate detection of particles in the ocean. However, the high sensitivity of the technology also makes it susceptible to interference from a large amount of information in the real environment[19-21]. The dyeing phenomenon of microplastics can also affect the accuracy of spectral identification. FTIR[22,23] is a technique that uses infrared light to measure the absorption and emission of different molecules. FTIR can provide information about the chemical composition and structure of microalgae, such as their lipid, carbohydrate, and protein content, FTIR also requires a higher concentration of the sample than other spectroscopic techniques. that can interfere with the infrared spectra of the sample. Therefore, FTIR measurements need to be performed in a dry and inert environment, which can increase the cost and complexity of the analysis. Fluorescence EEM combines different wavelengths of light to excite and measure the fluorescence emission of a sample such as microplastics and microalgae[24,25]. Fluorescence emission of microalgae can provide information about their biomass, composition, physiology, and stress response. Fluorescence EEM can also help to distinguish different species and strains of microalgae based on their spectral signatures. However, fluorescence EEM can be complicated by the overlapping or similarity of fluorescence spectra from different molecules or components in the sample, such as dissolved organic matter, microplastics, or other microorganisms. This can make it difficult to identify and quantify individual species or groups of species in the sample. Therefore, fluorescence EEM needs to be combined with other methods or techniques to improve its accuracy and reliability. Polarization describes how the electric field vector of a wave oscillates in a plane perpendicular to the direction of propagation. Polarized light scattering measurement is a method that uses polarized light to measure the scattering properties, such as their size, shape, orientation, and refractive index[26,27]. Polarized light scattering measurement can be used to detect and characterize SSPs in water, especially microplastics, which have different polarization responses than natural particles[28].

The polarization state of light can be described by a set of values called the Stokes vector. It consists of four parameters: *I, Q, U* and *V*, where *I* is the total intensity of light, *Q* is the intensity difference between the horizontal and vertical linear polarizations, *U* is the intensity difference between the +45° and -45° linear polarizations and V is the intensity difference between the right and left circular polarizations. The Stokes vector can be written as Eq 1. Alternatively, the polarization state can be represented by a point on a unit sphere, known as the Poincaré sphere. To determine the effect of an optical system on the polarization of light, one can use Mueller matrix to transform the Stokes vector of the input light into the Stokes vector of the output light, which can be written as Eq 2.

$$S = \begin{bmatrix} I \\ Q \\ U \\ V \end{bmatrix}, \tag{1}$$

$$S_{out} = \begin{bmatrix} I_{out} \\ Q_{out} \\ U_{out} \\ V_{out} \end{bmatrix} = \begin{bmatrix} m_{11}, m_{12}, m_{13}, m_{14} \\ m_{21}, m_{22}, m_{23}, m_{24} \\ m_{31}, m_{32}, m_{33}, m_{34} \\ m_{41}, m_{42}, m_{43}, m_{44} \end{bmatrix} \begin{bmatrix} I_{in} \\ Q_{in} \\ U_{in} \\ V_{in} \end{bmatrix}$$

(2)

The Poincaré sphere is then obtained by mapping the Stokes vector to a point on a unit sphere, using the following rules:

1. The I parameter is ignored, since it only affects the intensity of light, not its polarization state.
2. The Q, U and V parameters are used as the Cartesian coordinates of the point on the sphere, normalized by dividing by I.
3. The distance between two points on the sphere indicates how similar their polarization states are[29].

In this article, we present a novel method that combines multi-wavelength polarized light scattering and fluorescence EEM at 120° scattering angle to characterize the SSPs in water. The advantage of this method is that it can measure 27-dimensional independent optical parameters of SSP within 0.25μs, which is much faster and more accurate than existing methods.

We used 14 different kinds of SSP samples, which included seven harmful microalgae and common microplastics in marine pollution. Through simulation, we found that incident light with different wavelengths of the same polarization state can be well classified for the same SSP, which confirms the feasibility of the experimental device, and the results are promising.

Then, use MLP to train the network for classification with the simultaneous measurement of 27 parameters which contains 11 fluorescence signals and 16 scattering signals. And the experiments have found that these harmful microalgae can be well classified from each other. This work demonstrates the potential and application prospect of precise detection of SSPs in water environment.

## 2   Methods

### 2.1   Experimental setup

We show the basic optical path in Fig. 1, which consists of an illuminating optical path, a receiving optical path, and a custom-made beaker that holds the sample. The illuminating optical path is composed of a multi-color polarized light source that emits four wavelengths with 90° linear polarization, a polarization state generator, a diaphragm, and an achromat. Notably, the light source supports two modes: CW mode and modulated pulse light mode (MPL mode). Polarization state generator is consisting of a rotatable half wave-plate and a rotatable quarter wave-plate in sequence. The half wave-plate is used to convert the vertical linear polarization state of light source emitted into a linear polarization state in any direction, and a quarter wave- plate converts the linear polarization state converted by half wave-plate into the any polarization state[30]. Optical focusing is important for the adjustment of the illuminating optical path. In this part, we use a diaphragm to control the shape of the incident light. This ensures that the spots of four wavelengths on the achromat1 are consistent with each other. The diaphragm is essential for the optical focusing. We also consider the local tolerances of the lens to assess the impact of different wavelengths on the optical focusing. Eq 1 shows the local tolerance of the lens for different wavelengths, where f is the focal length of the achromat, 80mm, a is the diameter of the diaphragm, 10mm, and λ is the wavelength of the light source. The difference between the focus distances of the 445nm and 638nm lights through the lens is about 5μm, which is an acceptable error.

$$\Delta Z = \pm 3.2 \frac{\lambda}{2\pi}\left(\frac{f}{a}\right)^2 \sim \pm \frac{1}{2}\left(\frac{f}{a}\right)^2 \lambda \tag{3}$$

Sample is placed in a custom-made glass beaker, there is a magnetic stirrer in the beaker with a speed of 0-300 rounds per minute, which regarded as a simulation of underwater turbulence: the SSP in the sample pool rotating just like SSP keep moving in the force of turbulence in water environment. When the concentration of SSP in the sample pool is less than $10^5$ per milliliter, only one particle can be considered in the scattering volume statically.

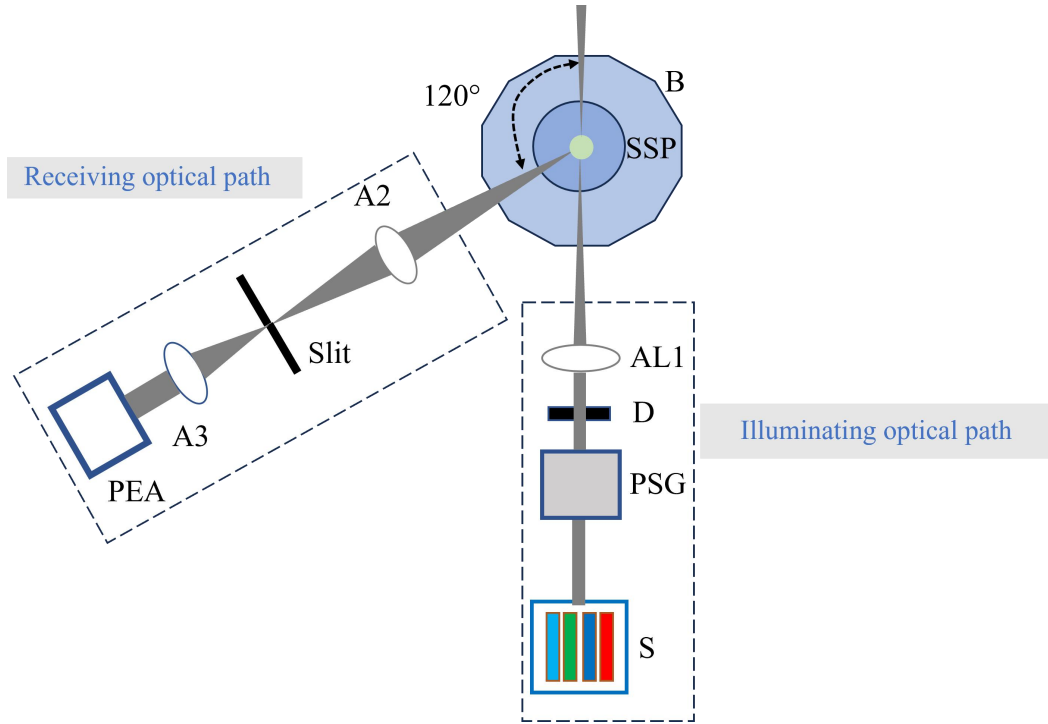

Figure 1 Optical setup scheme. S: Light source; PSG: Polarization state generator; D: diaphragm; A1: Achromat 1; SSP: Suspended single particle; A2: Achromat 2; A3: Achromat 3; PEA: Polarization State and EEM analyzer; B: custom-made beaker

The receiving optical path consists of two achromats, a slit, and a polarization state and EEM analyzer (PEA). The achromat 2 (A2) is carefully placed to collect the scattering and fluorescence of SSP. The SSP's location and the slit's location are at the conjugate planes of the slit, which can be achieved by adjusting A2. In addition, A2 and achromat 3 (A3), which have focal lengths of 30mm and 40mm respectively, form a spatial filtering device with a pinhole. This device limits the scattering volume to only 0.01μL. Then, A3 is adjusted to make the light perpendicular to its optical axis.

The PEA is a crucial device in the receiving optical path, as it can obtain the multi-wavelength Stokes vector and fluorescence EEM of SSP simultaneously. Therefore, it needs to be specially designed and calibrated, which enables us to convert the received optical information into Stokes vector and EEM directly in subsequent experiments. The PEA is connected to the photomultiplier tube (PMT), which in turn is connected to the parallel acquisition card (PAC). Firstly, we need to figure out the voltage-signal relation and then obtain the relevant transition equation, then using white light to put through the incident part of PEA and collect three channels' voltage of EEM analyzer to make sure whether the EEM analyzer can work normally. As shown in Fig. S3, through adding polarizers, quarter-wave plate and optical filters on beam splitters, we can realize these potential functions with low-cost, fastness and high accuracy in contrast to conventional polarization measurement. Secondly, we need to calibrate polarization analyzer and obtain calibration matrix C based on Sector 3 in Supplementary.

## 2.2 Signal Processing

There are two modes of light source, among which, the signal processing method in CW mode of SSP is the same as this paper[26]. It's noteworthy that the frequency of the modulated pulsed light (MPL) mode is set as 40KHz, which means the time interval of pulsed light of different wavelengths is much smaller than SSP appearing time, resulting in the scattered light of SSP in MPL mode is sophisticated, thus the signal processing in MPL mode is mainly introduced here. The time interval of pulsed light of different wavelengths ensures that polarization in different wavelengths and EEM of SSP can be collected by PEA simultaneously when the SSP is present in the detection region. Based on this, the 445nm light source is set to output a constant voltage as a monitoring signal when the 445nm light source emits. Firstly, SSP's scattered light's polarization of four wavelengths and EEM are extracted based on monitoring signal and the frequency of the MPL, as shown in Figure 2.

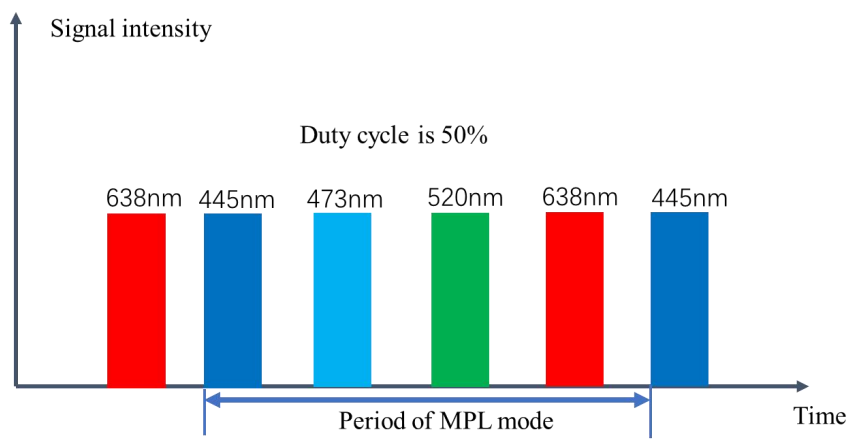

Figure 2 Schematic diagram of light source's modulated light mode

Then filtering the signal which reduces the impact of noise which comes from the environment, the scattering from water, interference caused by PAC and PMTs. Using the Butterworth low-pass filter which the cutoff frequency is 10KHz and the sampling frequency is 10MHz. We used the fundamental deep learning method for classification, multilayer perceptron. A multilayer perceptron (MLP) is a type of artificial neural network that consists of multiple layers of nodes, which can widely use for learning non-linear models from original data by adjusting the weights using a supervised learning method called backpropagation. MLPs are generally used for pattern recognition such as regression, classification, and feature extraction.

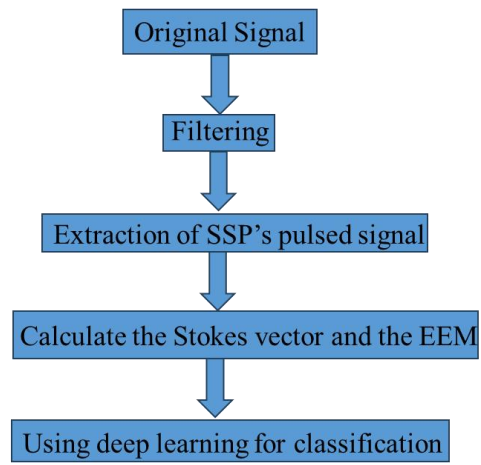

Fig 3 Data processing scheme

### 2.3   Samples

Different SSP containing microalgae, microplastics and sediments are used for classification. The samples include 10μm polystyrene microsphere (PS10), 10μm polystyrene microsphere with pores (P-PS10), *Cyclotella meneghiniana* (CM), *Chlorella* (CH), *Synechococcus elongatus* (SE), *Euglena gracilis* (EG), *Haematococcus Pluvialis* (HP), *Diatomite* (DI), *Dunaliella salina* (DS), *Chaetoceros tenuis* (CT), *Chaetoceros gracilis* (CG), *Thalassiosira weissflogii* (TW), *Skeletonema costatum* (SC), *Phaeocystis globosa Scherffel* (PG), *Nitzschia closterium f.minutissima* (NCF). The detailed information of samples in the work are shown in Table 1.The experimental samples we purchased from Nanomicro.Inc, Freshwater Algae Culture Collection at the Institute of Hydrobiology (FACHB) and Leadingtech.Inc.

Table 1

| SSP | Acronym | Species | Genus | Paticle size | Morphology | Sample's origin |
|---|---|---|---|---|---|---|
| 10μm polystyrene microsphere | PS10 | Microplastics | / | 10 microns | Spherical | Suzhou Nanomicro .Inc |
| 10μm polystyrene microsphere with pores | P-PS10 | Microplastics | / | 10 microns | Spherical | / |
| Cyclotella meneghiniana | CM | Diatom | Cyclotella | 6-18 microns | Spherical | FACHB |
| Chlorella | CH | Microalgae | Chlorophyta | 2-10 microns | Spherical | FACHB |
| Synechococcus elongatus | SE | Microalgae | Synechococcus | average length:3.8 microns average width:1.2 microns | Rod-shaped | FACHB |
| Euglena gracilis | EG | Microalgae | Euglena | Changing (20 microns-100 microns) | Unfixed | FACHB |
| Haematococcus Pluvialis | HP | Microalgae | Haematococcus | 30 microns | Spherical | FACHB |
| Diatomite | DI | Sediment | / | / | Unfixed | / |
| Dunaliella salina | DS | Microalgae | Dunaliella | Length:5-25 microns width: 3-13 | Pear shape | FACHB |

| | | | | microns | | |
|---|---|---|---|---|---|---|
| Chaetoceros tenuis | CT | Microalgae | Chaetoceros | 10 microns-50microns | elliptical or circular | LeadingTec |
| Chaetoceros gracilis | CG | Microalgae | Diatoms | / | Cylindrical | FACHB |
| Thalassiosira weissflogii | TW | Microalgae | Thalassiosira | 10 microns-30 microns | short cylinder | Leading Tec |
| Skeletonema costatum | SC | Microalgae | Skeletonema | 8 microns-12 microns | Cylindrical | LeadingTec |
| Phaeocystis globosa Scherffel | PG | Microalgae | Phaeocystis | 5 microns-10 microns | Spherical | FACHB |
| Nitzschia closterium f.minutissima | NCF | Microalgae | Nitzschia | 12 ~ 23 microns in length and 2-3 μm in width | Fusiform | FACHB |

## 3 Experimental Results

### 3.1 Contrast experiment

We can obtain from the simulation section in the supplementary materials that different wavelengths and polarization states can be used as degrees of freedom. This means that, theoretically, under the MPL mode of the light source, the information obtained by combining the multi-wavelength polarization and EEM of SSP is abundant. In this section, we will explore the details of this type of enhancement. As mentioned above, we successfully obtained the polarization of multi-wavelength and EEM of SSP, and then used MLP to classify seven kinds of SSP. The input features of MLP are the multi-wavelength Stokes vector and EEM. Fig 4 shows the experimental results. We can see that all chlorella illuminated by 520nm laser is misjudged as Meniella, and a high proportion of polychlorella slender is also misjudged as Meniella.

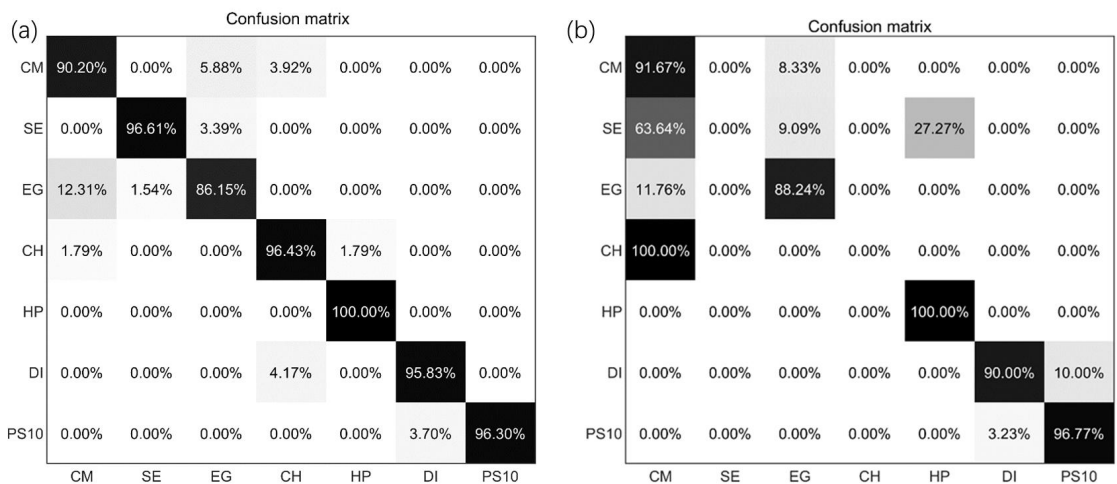

Figure 4 Contrast experiments (a) Classification results of seven kinds of SSP illuminated by light source's MPL mode; (b) Classification results of seven kinds of SSP illuminated by 520nm laser

The contrast experiment proves that the device has good classification performance for SSP which is different in the surface roughness, refractive index and size. It is evident that obtaining multi wavelength polarization information of SSP can predominantly improve the accuracy.

### 3.2　Seven harmful microalgae classification

Algae blooms are common phenomena in marine environments, which involve the rapid growth of certain microalgae species. These species include Chaetoceros tenuis (CT), Dunaliella salina (DS), Nitzschia closterium f. minutissima (NCFP), Paeocystis globosa Scherffel (PGS), Skeletonema costatum (SC), Thalassiosira weissflogii (TW), and Chaetoceros gracilis (CG). Algae blooms can have negative impacts on the marine ecosystem and human health, as they can deplete oxygen, nutrients, and other substances in the water, resulting in eutrophication.

|     | DS     | CT     | CG    | TW    | SC      | PG     | NCF     |
|-----|--------|--------|-------|-------|---------|--------|---------|
| DS  | 84.31% | 0.00%  | 7.84% | 7.84% | 0.00%   | 0.00%  | 0.00%   |
| CT  | 0.00%  | 90.67% | 0.00% | 9.33% | 0.00%   | 0.00%  | 0.00%   |
| CG  | 7.25%  | 0.00%  | 86.96%| 1.45% | 0.00%   | 4.35%  | 0.00%   |
| TW  | 0.00%  | 2.27%  | 0.00% | 97.73%| 0.00%   | 0.00%  | 0.00%   |
| SC  | 0.00%  | 0.00%  | 0.00% | 0.00% | 100.00% | 0.00%  | 0.00%   |
| PG  | 0.00%  | 0.00%  | 0.00% | 0.00% | 0.00%   | 94.59% | 5.41%   |
| NCF | 0.00%  | 0.00%  | 0.00% | 0.00% | 0.00%   | 0.00%  | 100.00% |

Figure 5 Classification results of seven kinds of harmful microalgae

Some algae blooms also produce toxins that can harm aquatic organisms and humans, such as CT and DS. Moreover, algae blooms can reduce the light penetration in the water column, hindering the photosynthesis of underwater plants and algae. In severe cases, algae blooms can lead to the collapse of underwater ecosystems. As shown in Figure 5, the device was applied to the classification of seven types of bloom algae, and the classification accuracy exceeded 93% which shows that the device can distinguish harmful algae in marine.

## 4    Conclusions and discussions

Simulation provides solid theoretical foundation for experiments. In the sector 4 of supplementary, we can see the diverse results of the suspended single sphere of different refractive index and size illuminated by the light of different wavelength and polarizations at a 120-degree scattering angle. The experimental results confirm the feasibility of combining multi-wavelength polarized light scattering with EEM for the detection of single particle. Polarized light scattering can provide information about the microstructure, size, and shape of particle, whereas EEM can characterize the cell's chemical composition. The collaboration guaranteed by synchroneity gives us comprehensive information for the deeper exploration of the physical and chemical property of SSP. The effectiveness of simultaneous detection has been strictly tested in well-designed experiments.

We have developed a device that can simultaneously measure the multi-wavelength polarized light scattering and EEM of SSP, which are independent to each other. This device can obtain 27-dimensional effective and original optical information of SSP at once, which is strictly tested and can be used for particle classification with better results than traditional methods. To achieve this goal, we have designed a light source that can generate the monitoring signal in MPL mode, and an algorithm for SSP's signal extraction. We can successfully obtain the effective information of SSP by using these methods. We have also creatively introduced the time dimension to extract particle's information at the microsecond scale. This is a novel approach, as previous research on particulate matter in water has not used polarized light scattering and EEM at the tiny time scale to characterize SSP. By introducing the time dimension, we can collect more abundant information than traditional research, providing rich physical and chemical information for SSP's classification and recognition.

## 5 Author Contributions

RL designed the experimental device and participated in the design of the light source. TL built the equipment construction, including the construction of the settlement, optical path calibration, digital circuit connection, code design. MY participated in the code writing, especially in classification algorism and signal process, JY participated in the classification algorism, ZX participated in the design and manufacture of the splitter cage, and Ma Hui provided the setup for the overall experiment

## 6 Acknowledgments

We thank assistance in the optical source's manufacture by the staff at Suzhou Institute of Biomedical Engineering and Technology of Chinese Academy of Sciences. We thank the staff at Guangdong Research Center of Polarization Imaging and Measurement Engineering Technology.

## 7  Sector 1 Light source

Multi wavelength light sources have two modes: continuous mode (CW mode), where one or more wavelengths can be selected simultaneously, and modulated pulse mode, where four wavelengths can be selected for output, with adjustable duty cycle and pulse frequency. The main components of the light source are the optical power, frequency, duty cycle, and illumination conditions (MPL mode or CW mode) for each of the four-wavelength modulated light types. To control the output mode of the laser, the user needs to input the parameter values and press the send data button, which transmits the data to the laser. The 445nm laser is used to monitor the MPL mode of the light source, which allows us to infer the occurrence of the other three lasers based on the frequency and duty cycle in the setting menu (see Fig S1).

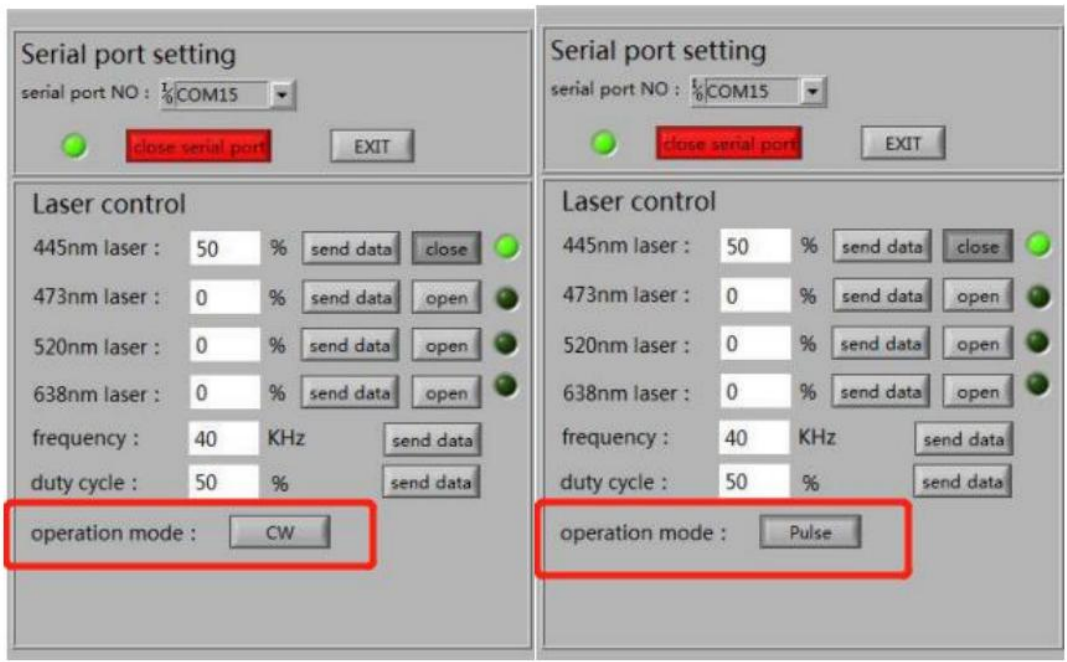

Figure S1 Setting menu of light source

The horizontal and vertical adjustment of the light from source is crucial for the setup that guarantees the effectiveness of the received information in the process of adjusting the optical path. The light source consists of four lasers with wavelengths of 445 nm, 473 nm, 520 nm, and 638 nm. It was found that the beam center of the four-wavelength light source was horizontal and the edges were divergent. In order to make the light spots of the four wavelengths incident on the lens have the same size, resulting in better achromatic and focusing effects for multi-wavelengths. A diaphragm can be used to limit the size of the light spot, remove the influence of its edge, and allow the central part of each wavelength to be illuminated on the sample through an achromatic lens. By adjusting the holder of the light source and adding a diaphragm, we can align it horizontally and vertically. Then we tested this light source, which is specially customized, to ensure that it can illuminate the lens L1 normally with multi-wavelengths, as shown in Figure S2.

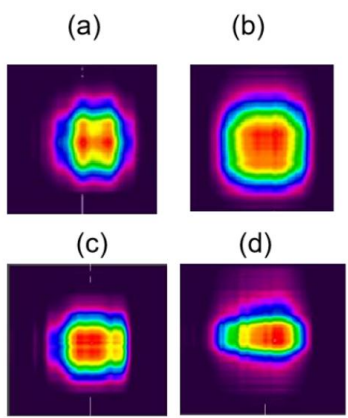

Figure S2 Spot contour away from the light source outlet (a) 445nm laser's spot contour; (b) 473nm laser's spot contour; (c) 520nm laser's spot contour; (d) 638nm laser's spot contour

## 8 Sector 2 PEA's design and Calibration

The design of a beam splitter cage can better distribute light energy, since the fluorescence signal intensity of biological particles is weaker than polarization. The polarization scattering of particles is generally stronger than the fluorescence signal, which is worth noting. As a result, we tested two different beam splitters of BS4 with ratios of 3:7 and 1:9, as shown in Figure 2 and Figure S3, and ultimately selected a beam splitter prism with a 3:7 ratio for use in PEA, that is to say, 70% of received light is distributed to calculate the EEM. The EEM module separates the spectral signal of fluorescence through a combination of splitter prisms and optical filters.

Foreman et al. showed that the polarization state of light can be completely determined by at least four distinct projective measurements of the corresponding Stokes vector[1]. Therefore, we designed four independent polarization detection channels by adding linear polarizers and quarter-wave plates. To obtain SSP's polarization, we need to calibrate the output optical path, including all the optical components by means of well-designed method mentioned in the Sector 3. The purpose of calibration is to convert channel information into Stokes vector information accurately through obtaining the calibration matrix, and the focus of acquired matrix is the condition number, which is used to measure how sensitive Stokes vector is to changes or errors in the input, and how much error in the output results from an error in the input.

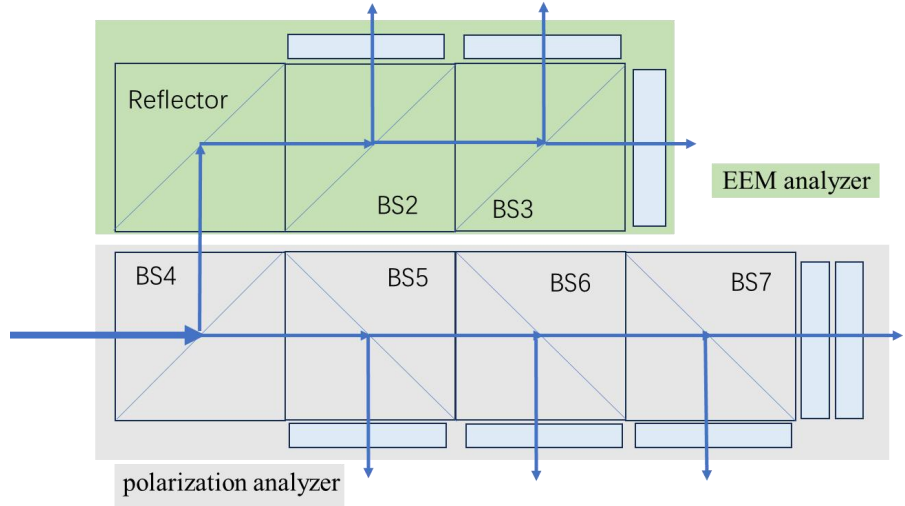

Figure S3 PEA scheme. BS: beam splitter cube; F1: 638nm band-pass filter, band width: 30nm; F2: 685nm band-pass filter, band width: 30nm; F3: 710nm long-pass filter; P1: 0-degree polarizer; P2: 45-degree polarizer; P3: 90-degree polarizer; P4: 135-degree polarizer; QW: quarter- wave plate.

## 9 Sector 3 Calibration of optical path

As we have discussed, the PEA needs to be calibrated to obtain reliable polarization information and EEM with good performance. Therefore, we will illustrate the calibration process in this section.

We used the following instruments for our experiment: a collimated light source which emit high-parallel light rays, an attenuation plate, a half-wave plate, a quarter-wave plate, PEA, PAX1000VIS (purchased from Thorlabs.inc), several PMTs, an acquisition card, and so on.

We use polarization state generator (PSG), which is same as incident light path, to modulate the appropriate number of polarization states of incident light for the entire calibration process. As shown in figure S4. Firstly, utilizing PSG to modulate several kinds of polarization, then using PAX1000VIS to record the relevant polarization, in other words, obtaining a matrix $A$ containing several Stokes vectors, afterwards, remove the PAX1000VIS and using the same operation to modulate the incident light, then collect the receiving four-channel information of PEA, recorded as matrix B, then by means of matrix calculation, we ultimately obtained the calibration matrix, as shown in Equation S1.

$$C = A * inv(B) \qquad \text{S1}$$

The condition number of calibration matrix is 3.8, conventionally, we assume that condition numbers which below 4 are reasonable, that is to say, render the measurement error varying from a rational range. The performance of PEA is simple and the method is inexpensive, scalable, and high-sensitivity for micrometer-scale objects. Through the study mentioned above, fixed size devices can be customized and installed in different application devices to characterize SSPs with relatively weak scattering.

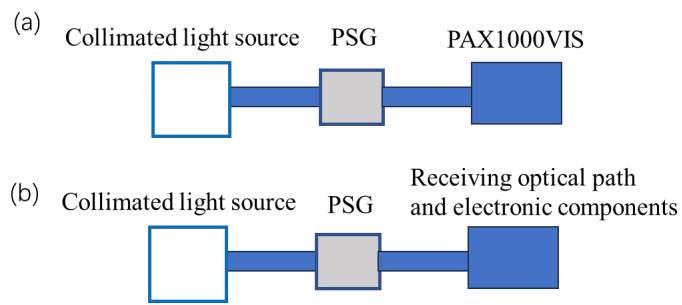

Figure S4 Calibration process. Step (a) Polarization measurement of incident light; Step (b) Calculation of calibration matrix.

**10   Sector 4 Simulation of Mie particles**

Mie scattering simulation provides guidance and assistance for experiments. Figure S5- Figure S11 show the simulation results of SSPs' Stokes vector located at 120° scattering angle. By the way, λ1, λ2, λ3 and λ4 in these figures refer to 445nm, 473nm, 520nm and 638nm of light source, individually. Typically, we used normalized Stokes vector in the simulation and experiment.

Two kinds of simulations are provided here: the first one is the Stokes vector of SSPs with a size of 10μm and refractive index ranging from 1.3 to 1.8 under the irradiation of four wavelengths. The reason for selecting the setup is:

1) The microplastic spheres with a diameter of 10μm are made in a standard way and widely used in experiments;

2）The refractive index, ranging from 1.3 to 1.8, is the common refractive index range for SSPs in water environment.

Figure S5 shows the SSP's scattering illuminated by the 135-degree linear polarized light.

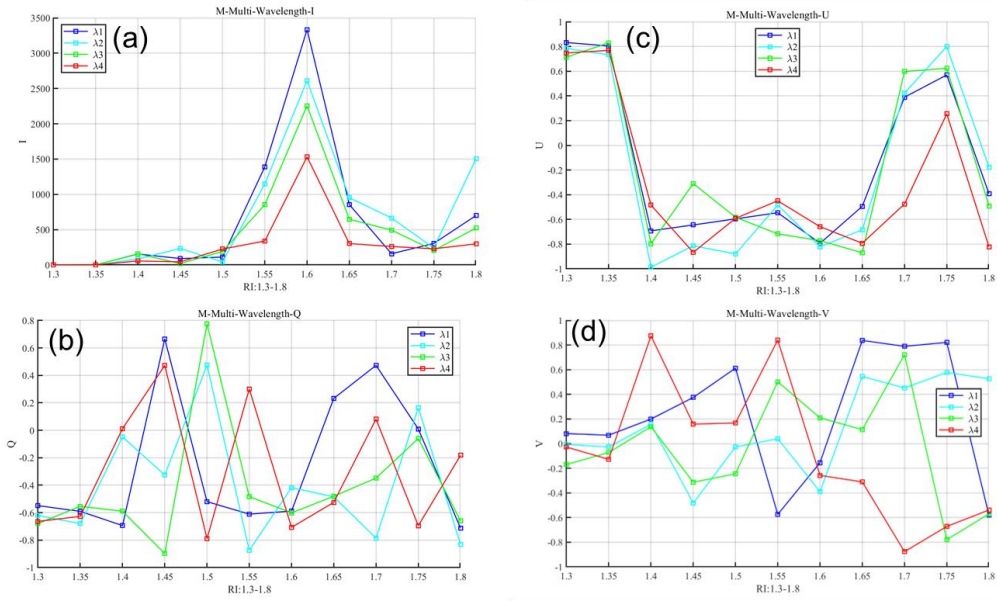

Fig S5 SSPs illuminated by 135-degree linear polarized light (particle size: 10μm, refractive index: 1.3-1.8) (a) I; (b) Q; (c) U; (d) V.

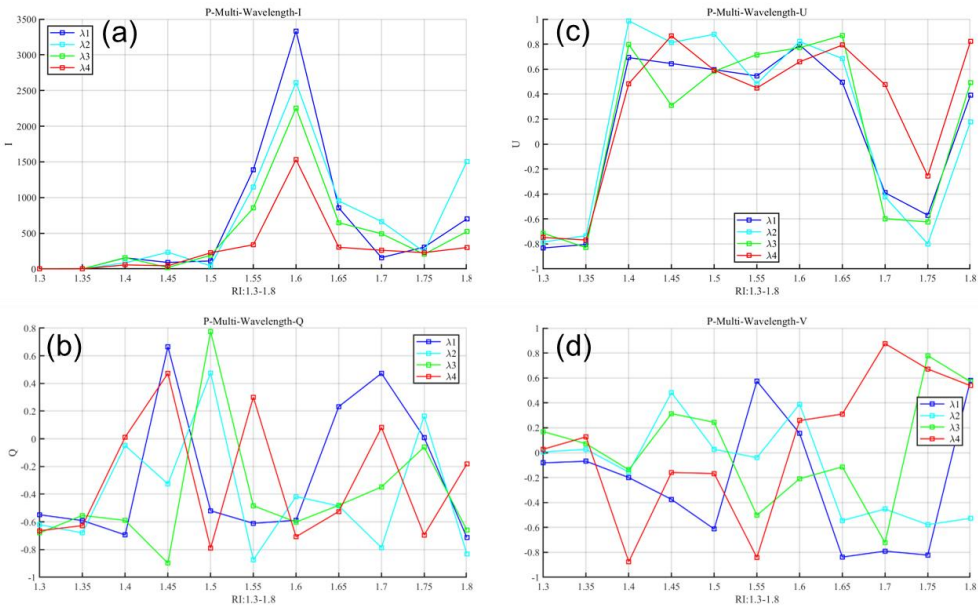

Fig S6 SSPs illuminated by 45-degree linear polarized light (particle size: 10μm, refractive index: 1.3-1.8) (a) I; (b) Q; (c) U; (d) V.

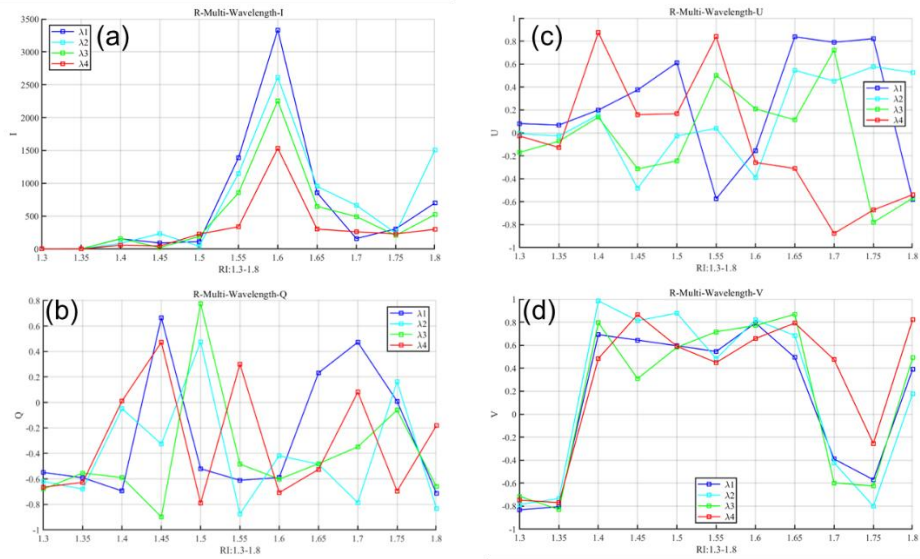

Fig S7 SSPs illuminated by right circular polarized light (particle size: 10μm, refractive index: 1.3-1.8) (a) I; (b) Q; (c) U; (d) V.

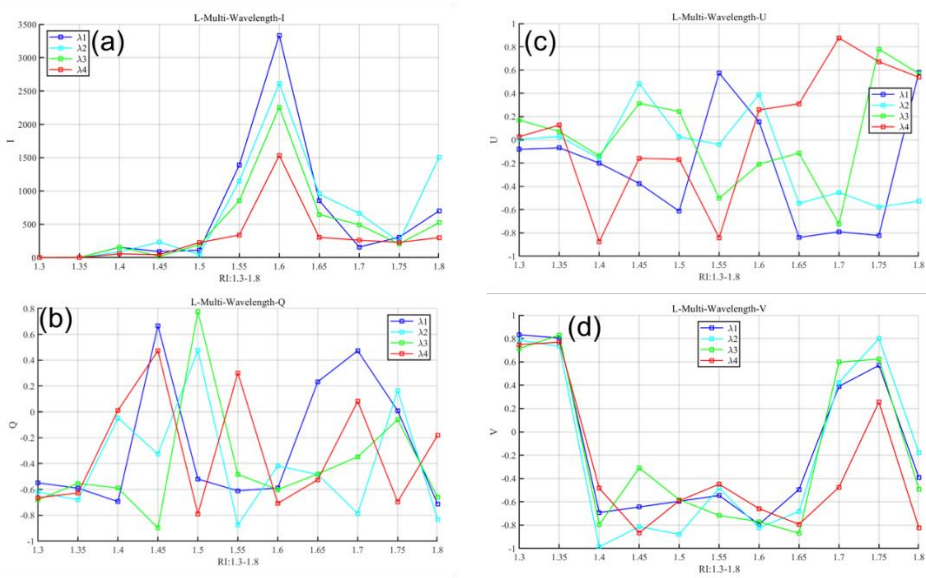

Fig S8 SSPs illuminated by left circular polarized light (particle size: 10μm, refractive index: 1.3-1.8) (a) I; (b) Q; (c) U; (d) V.

The simulation results above shed light that light intensity is the highest among all refractive indices when the refractive index is 1.6, which provides us with important reference information for choosing appropriate samples in experiments. In order to explore the valuable information, the following simulation uses SSPs with refractive index of 1.6 and particle size ranging from 2μm to 60μm as the simulation conditions.

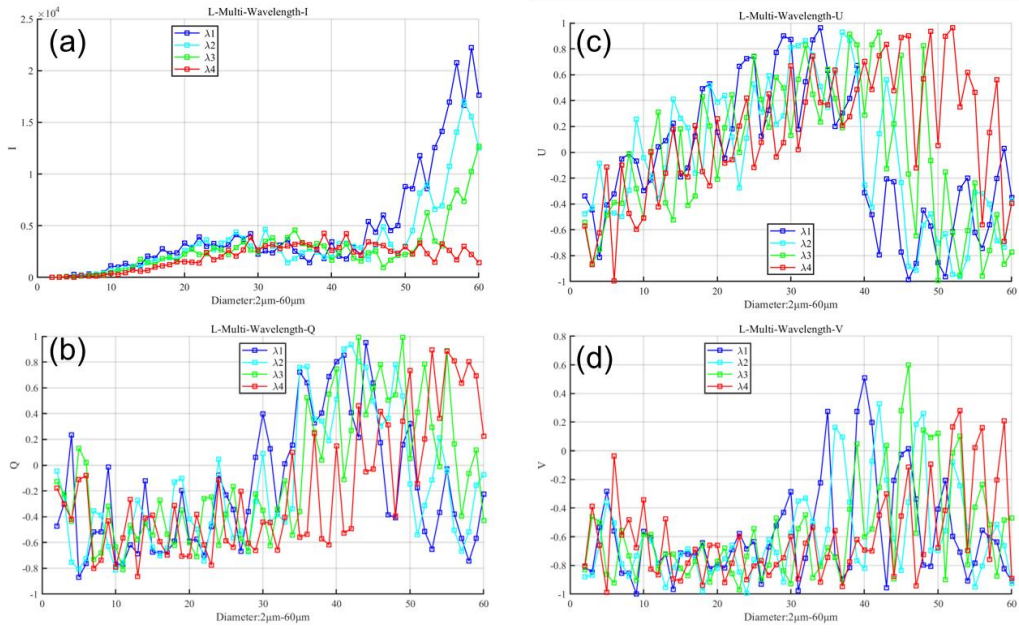

Fig S9 SSPs illuminated by left circular polarized light (particle size: 2-60μm, refractive index: 1.6) (a) I; (b) Q; (c) U; (d) V.

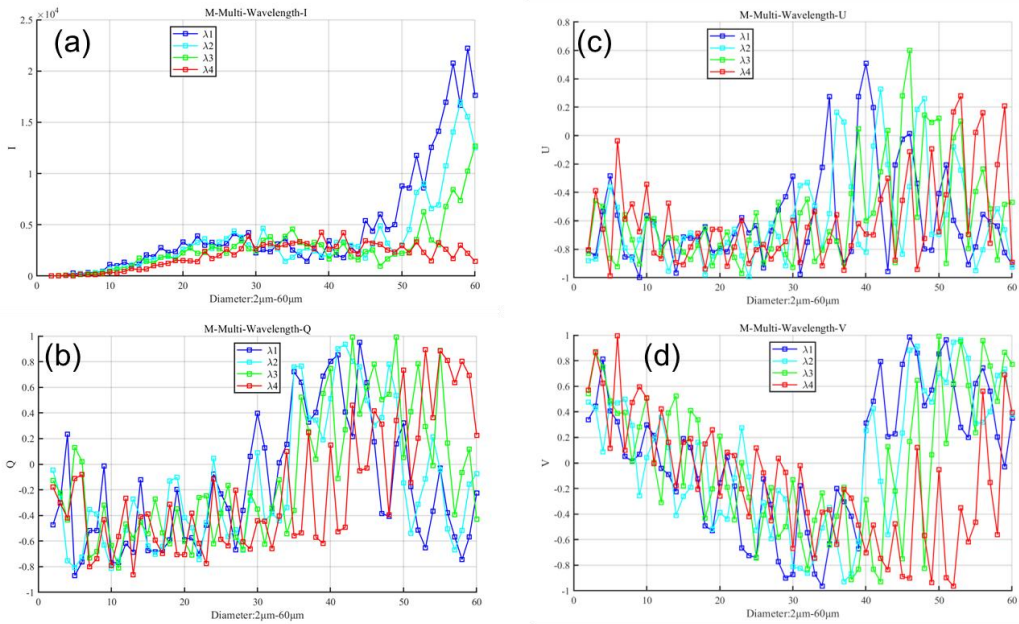

Fig S10 SSPs illuminated by 135-degree linear polarized light (particle size: 2-60μm, refractive index: 1.6) (a) I; (b) Q; (c) U; (d) V.

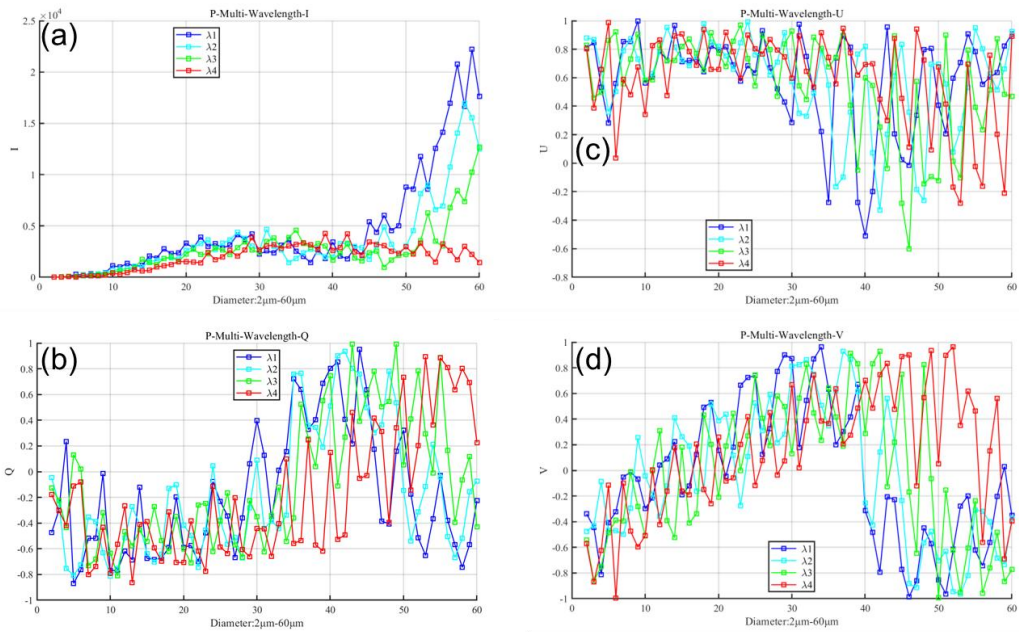

Fig S11 SSPs illuminated by 45-degree linear polarized light (particle size: 2-60μm, refractive index: 1.6) (a) I; (b) Q; (c) U; (d) V.

The results above indicate that the polarized light scattering of the same particle is different under different wavelength and polarizations, we will see that the experimental results are basically consistent with the simulation results, that is to say, by adding the lasers with different wavelength is equivalent to adding a degree of freedom that has strictly proved.

1    Foreman, M. R., Favaro, A. & Aiello, A. Optimal Frames for Polarization State Reconstruction. *Physical Review Letters* **115**, 263901, doi:10.1103/PhysRevLett.115.263901 (2015).